~~~

# Comment on "The influence of planetary attractions on the solar tachocline" by Callebaut, de Jager and Duhau

N. Scafetta[a*], O. Humlum[b,c], J.-E. Solheim[d], K. Stordahl[e]

[a] ACRIM (Active Cavity Radiometer Solar Irradiance Monitor Lab) & Duke University, Durham, NC 27708, USA

[b] Department of Geosciences, University of Oslo, Oslo, Norway

[c] Department of Geology, University Centre in Svalbard (UNIS), Svalbard, Norway

[d] Department of Physics and Technology, University of Tromsø, Tromsø, Norway

[e] Stovnerbakken 10, 0980 Oslo, Norway

## Abstract:

Callebaut et al. (2012)'s claim that Scafetta (2010)'s results about a correlation between 20-year and 60-year temperature cycles and the orbital motion of Jupiter and Saturn were not confirmed by Humlum et al. (2011) is erroneous and severely misleading. Also Callebaut et al. (2012)'s *absolute* claim that a planetary influences on the Sun should be ruled out as a possible cause of solar variability is not conclusive because: (1) their calculations are based on simplistic classical Newtonian analytical mechanics that does not fully characterize solar physics; (2) the planetary theory of solar variation is supported by empirical findings. We show that both claims are already questioned in the scientific literature.

**Keywords:** theory of solar and climate variation; harmonic component of solar and climate variation; interpretation of solar and climatic proxy models.

[*]Corresponding author: nicola.scafetta@gmail.com

## 1. Introduction

Herein we correct some errors we find in Callebaut et al. (2012) about whether empirical results found in Scafetta (2010) were contradicted by Humlum et al. (2011). We show that no contradiction exists once that the two papers and their references are carefully read. In addition, other claims made in Callebaut et al. (2012) about a planetary influence on solar activity appear to be contradicted by past and recent published literature and may be philosophically problematic. Finally, we add a short Appendix to briefly respond to the "replay" to our comment (made in Sections 2 and 3 of this paper) by Callebaut, de Jager and Duhau (2013) hoping to make this exchange into a scientific important discussion.

## 2. Climate and astronomical cycles

About the first issue, in Callebaut et al. (2012) we read the following paragraph on page 74: *"Another

*approach to the problem is the study of climate variations in attempts to search for planetary influences. As an example, we mention a paper by Scafetta (2010), who found that climate variations of 0.1- 0.25 K with periods of 20-60 years seem to be correlated with orbital motion of Jupiter and Saturn. This was however, not confirmed in another paper on a similar topic (Humkin et al., 2011)."*

Callebaut et al. (2012) criticized Scafetta's work in the above brief reference by providing a single contrary reference without any discussion. However, the cited reference (*Humkin, O., Solhelm, J.-E., Stordahl, K., 2011. Identifying natural contributions to the late Holocene climate change. Global and Planetary Change 78, 145–156*), which would not confirm the findings in Scafetta (2010), does not exist in the scientific literature.

Apparently, there are four serious typos in the reference, and Callebaut et al. (2012)'s intention was to refer to: *Humlum, O., Solheim, J.-E., Stordahl, K., 2011. Identifying natural contributions to late Holocene climate change. Global and Planetary Change 79, 145–156.*

Humlum, Solheim and Stordahl would like to point out that in their paper they do not at all discuss correlations of climatic cycles with planetary orbital periods as done in Scafetta (2010). The primary aim with their investigations was to search for possible periodic temperature variations at the secular/millennial scale in a multi-millennial proxy temperature record and use those oscillations for prediction of temperature in the near future. Humlum et al. (2011) analyzed a 4000 yr long ice core series (the GISP2 record from central Greenland; Alley, 2000; which is a low resolution record), and found significant periods at about 550 and 1100 yr, which may explain the present warming since 1700 as part of a millennial cycle. Moreover, proxy temperature models are affected by large errors (Bender et al., 1997) and describe only *hypothetical* temperature histories, and their accuracy in correctly reproducing the temperature details should not be overstated. This is also why Humlum et al. (2011) only used 3 periods in the range 500-2800 yrs in their predictions. These temporal scales are far larger than the decadal/multidecadal scales studied in Scafetta (2010), and no direct comparison between the two studies can be made.

In addition, Humlum et al. (2011; 2012) analyzed a 98-year long temperature series from Svalbard (1912-2010), a small archipelago in the Arctic. This is a local meteorological temperature record, it is significantly shorter than the global temperature records used in Scafetta (1850-2010) and it does not need to exactly reproduce the patterns observed in the global temperature records. In any case, in the latter record, Humlum et al. (2011; 2012) found a dominant modulation of the order 60-70 years. Also temperature series from a number of meteorological stations in Norway and in the North Atlantic region, in most cases back to 1860, analysed by Solheim et al (2012) all showed temperature maxima around the 1880s, 1940s and 2000s. This pattern may well correspond to Scafetta's quasi 60-year cycle found in the global surface temperature records from 1850 to 2010 that correlates well with the quasi 60-year Jupiter/Saturn harmonic. In fact, Solheim et al (2012) suggested (p. 282) that the solar cycle length (SCL) is related, in some way, to astronomical forcing.

Thus, the statement found in Callebaut et al. (2012) is a misunderstanding of Humlum, Solheim and Stordahl's paper. It is not correct to claim that the correlation between the global surface temperature 20 and 60 yr cycles and 20 and 60 yr planetary harmonics, [which is so clearly shown in figures 10 and 11 in Scafetta (2010), in figure 10 in Scafetta (2012a) and again in Scafetta (2012b) during the analyzed period 1850-2010] were contradicted in Humlum et al. (2011), or by other studies authored by the same authors.

Scafetta would like to add the following. Scafetta (2010) studied 9 global surface temperature records from 1850 to 2010 (the nine records are Global, Land, Ocean, in 3 versions: global, North and South) and looked for major decadal and multi-decadal cycles. Scafetta found synchronicity among the nine analyzed global temperature records and between them and a set of astronomically deduced cycles. About the 10, 20 and 60-year cycles Scafetta supported his results by referencing numerous other studies that Callebaut et al. (2012) have not taken into consideration.

Indeed, scientific literature establishes that Scafetta's 20 yr and 60 year cycles are present in Greenland ice core records and in the Arctic as well as in other climatic long records. For example, the harmonics present in the GISP2 record studied in Humlum et al. (2011) were also analyzed, using a longer period, in Davis and Bohling (2001, figure 7) and in Knudsen et al. (2011, figure 5 c-d), who found a significant power spectrum peak at about 60-year period among other cycles. Klyashtorin and Lyubushin (2007) and Klyashtorin et al. (2009) in their analysis of changes of fish productivity in the Barents Sea observed a cyclic variation with an about 60 year period: these authors also observed a similar periodicity in numerous multisecular climatic records. Also global records of sea level rise (Jevrejeva et al., 2008), of North Atlantic Oscillation index (Mazzarella and Scafetta, 2012), of the Atlantic Multidecadal Oscillation (AMO) and Pacific Decadal Oscillation (PDO) (Loehle and Scafetta, 2011) present a 60-year major modulation for centuries. The existence of a quasi 60-year warming/cooling cycle from 1940 to 2000 has recently been confirmed even by a study on historical aerial photographic images of the Greenland southeast glaciers (Bjørk et al., 2012). In general, 50-70 year oscillations, together with other cycles are found in the climate system in numerous multisecular records (Schlesinger and Ramankutty, 1994; Klyashtorin et al., 2009; Qian and Lu, 2010; Loehle and Scafetta, 2011 and numerous references in it).

Figure 1 shows the GISP2 record used in Humlum et al. (2011) since 1350 AD (red), its power spectrum evaluation (in the insert) and the 60-year astronomical cycle proposed by Scafetta (2010) (blue). Despite the low resolution of the GISP2 ice core temperature proxy record, which deteriorates going back in time, during the analyzed period the proxy record clearly shows a major 60-61 year cyclical modulation, which is in relatively good phase with Scafetta's 60-year astronomical harmonic.

Thus, at least 10-11 consecutive 60-61 year cycles can be found in the GISP2 record plus the temperature records of the 20[th] century from 1350 to 2000, which include the two quasi 60-year cycles observed in the global surface temperature since 1850. A quasi 20-year cycle is also found for centuries and millennia in the Arctic (Chylek et al., 2011; 2012). About the 20-year cycle, note that it appears to beat because of the presence also of a climatic effect of the 18.6 yr lunar nutation cycle.

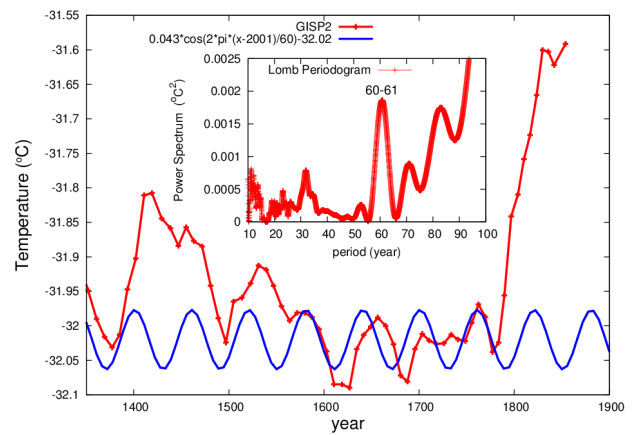

**Figure 1.** The GISP2 record in Humlum et al. (2011) from 1350 AD to 1850 AD (red), its power spectrum evaluation (in the insert) and the 60-year astronomical cycle proposed by Scafetta (2010) (blue). Despite a possible temporal and amplitude uncertainty, the GISP2 record presents a quasi 60-year modulation (clearly revealed in the power spectrum 60-61 yr peak) in relatively good phase with the 60-year astronomical harmonic proposed by Scafetta (2010). There appears to be a small time-lag perhaps due to a climatic heat capacity delay or to a possible proxy timing error (Bender et al., 1997).

## 3. Planetary and solar oscillations: empirical evidences and theory

About the major topic addressed in Callebaut et al. (2012) (the plausibility of a planetary influence on the Sun), these authors claim: (1) planetary oscillations are not able to reproduce the harmonics found in the solar activity records (this statement was provided without any discussion); and (2) various accelerations near or in the tachocline area are found to be larger than those due to the attraction by the planets by about three-four orders of magnitude. From the above, Callebaut et al. concluded that *"a planetary influences should be ruled out as a possible cause of solar variability"*.

A detailed response/rebuttal to both Callebaut et al. (2012)'s claims is contained in Scafetta (2012c, 2012d) and in the numerous references there cited. Therefore, we do not repeat the mathematical details and full arguments here, but invite the interested reader to read those papers.

Indeed, there are numerous empirical evidences of a planetary modulation of solar activity at multiple time scales (for example, see: Hung, 2007; and Scafetta, 2012a, 2012c, 2012d) and preliminary physical models are provided in Scafetta (2012d) and in Wolff and Patrone (2010). We refer to the above references for detailed discussions where empirical evidences for a planetary origin of multiple solar cycles (e.g.: sub-annual, Schwabe, Hale, Gleissberg, De Vries and Hallstadt cycles), solar activity hindcast models based on planetary harmonics and a quantitative physical model to calculate the amplification effects due to nuclear fusion rate feedback to tidal work have been proposed.

Moreover, Hanasoge et al. (2012) have recently found that the interior convection motions of the Sun appear to be up to hundred times weaker than those predicted by traditional theoretical models of solar convection similar to those used in Callebaut et al. (2012). This result too would greatly mitigate the claims in Callebaut et al. (2012) and in de Jager and Versteegh (2005), and question their conclusions.

Indeed, the phenomenon of a planetary influence on the Sun and/or on the climate appears to be more complex than what Callebaut, de Jager, and Duhau have hypothesized in their 2012 study. They based their argument on mere simplistic classical physics considerations, e.g. basic evaluations of theoretical tidal accelerations and Coriolis forces at just the tachocline. However, Callebaut et al. (2012) did not take into account that tidal forces act everywhere inside the Sun, also in the core where they would interfere with the luminosity production. So, there is a need to integrate the tidal effects at all distances from the solar center to the surface and evaluate the core response to the gravitational perturbations that the planetary tides cause in it, as Scafetta (2012d) assumed in his calculations.

It is worth to remind that simple Newtonian analytical mechanics by alone does not fully characterize stellar or solar physics, as it is well known. For example, in the 19$^{th}$ century Lord Kelvin used simple classical physics and concluded that: (1) the Sun had to be about 10-40 million years old (the Kelvin-Helmholtz Timescale; Carroll and Ostlie, 2007, page 296); and (2) the sunspots and geomagnetic activity could not be connected simply because of the too large Sun-Earth distance (Moldwin, 2008, page 11). Both claims were wrong.

As also acknowledged in de Jager and Versteegh (2005), solar dynamo theory by alone does not explain why the Schwabe cycle has an approximate average 11-year period nor does it explain the secular (Gleissberg) and multisecular (De Vries) oscillations producing, for example, the Maunder and Dalton like solar grand minima or other solar oscillations such as the millennial one (Hallstadt cycle). The solar dynamo theory just predicts the existence of a sunspot cycle, but it does not predict that for our Sun the length of this cycle must be around 10-12 years, which is a result that is roughly obtained only by choosing specific values for the free parameters of the dynamo models, which are directly measured on the Sun and finely tuned according to the specific model (Jiang et al, 2007). The timing of the 11-year solar cycle is also not predicted but constrained in the initial conditions of the dynamo models. On the contrary, other studies as well as Scafetta found numerous evidences for an astronomical signature in both climatic and solar records by using spectral coherence and synchronization analysis: facts that were not contradicted by Humlum et al. (2011) as erroneously claimed. The issue of whether and how the planets can influence or modulate solar activity should be considered open to scientific investigation also because an alternative theory that explains the observed solar oscillations does not exist.

## Appendix. A brief response to the reply by Callebaut, de Jager and Duhau (2013).

We thank Callebaut, de Jager and Duhau (2013) for their reply. They have acknowledged that: (1) they have not carefully proofread their paper; (2) their reference to Humlum et al. (2011) to criticize Scafetta (2010) was not correct nor appropriate, and they have withdrawn it; (3) *"there are some periodicities that are common to solar activity and planetary motions"*, which contradicts the main claim made in their original 2012 paper that *"none of the papers on planetary influences on solar variability succeeded in identifying"* major solar cycles as correlated to planetary cycles.

About the third point we observe that in addition to the previous literature on the planetary influence on solar and climate activity (properly referenced in Scafetta's papers), during the last months numerous additional papers have been published further demonstrating that major solar oscillations at multiple scales appear to be well correlated and/or synchronized to planetary motion (e.g.: Leal-Silva and Velasco Herrera, 2012; Abreu et al., 2012; Scafetta 2012c, 2012d; Tan and Cheng, 2013; Scafetta and Willson, 2013). See also the recent commentary by Charbonneau (2013) on Nature talking about a *"revival"* of the planetary hypothesis of solar variation.

About the last statement by Callebaut, de Jager and Duhau (2013) criticizing Scafetta (2012d) (quote from their reply: *"But, apart from some rough energy considerations he does not produce a sound physical analysis of the strength of the amplified signal to prove that it is "sufficiently energetic" to activate the proposed mechanism"*), it is incorrect because Scafetta does quantitatively evaluate the amplified signal energy and demonstrates that it is within one order of magnitude compatible with the total solar irradiance fluctuation. This order of accuracy is compatible with the physical accuracy required to demonstrate with Newtonian physics that the ocean tides on the Earth are regulated by the Moon and the Sun.

Not all physical details on this phenomenon are already known for sure. However, science gradually progresses from the mere observation and discovery of empirical correlations to a full microscopic physical understanding of a phenomenon. This process has not prevented people from fairly acknowledging the plausibility of scientific theories even if based only on empirical evidences. For example, since antiquity, ocean tides were linked to the lunar phases, efficiently forecasted and widely used to make choices and predictions in economics, farming, medicine, and navigation without any need of using the 18$^{th}$ century Newtonian gravitational theory, as Kepler (1601), who lived one hundred years before Newton, elegantly explained us.

The impression is that Callebaut, de Jager and Duhau (2013)'s argument is that a planetary theory of solar variation cannot be taken seriously if not already backed up with a solid and conclusive analytical physical explanation to be provided in the so-called *"well-founded paper"*. This argument is a repetition of a conservative, and likely erroneous way of reasoning that twists the normal process of science. Science progresses from an observational empirical theory of a natural phenomenon to a more mathematical based analytic theory, not vice-versa. Right now, the planetary theory of solar variation is not just a mere conjecture. It is based on a large number of observations, empirical hindcasting models, and preliminary physical models, as shown in the references that, evidently, need to be properly studied and evaluated by those who may not be familiar with this literature.

It is worth to remind that shortly before World War I the German scientist Alfred Wegener proposed a continental drift theory based on numerous, accurate and very convincing complementary empirical evidences based on geographical, sediment and fossil pattern matching across continents. However, his hypothesis was met with unfair skepticism from largely conservative scientists, who were resistant to any innovative geophysical theory, simply because Wegener did not have a definitive physical explanation about what made the continents move. The criticism was unfair because empirically based theories cannot be rejected simply because an analytical dynamical theory is still missing or may appear incomplete. Today Wegener's hypothesis is fully acknowledged, not because we now have a solid and definitive physical explanation (it is still missing), but simply because all investigations on sedimentation around the continents (including oil + gas exploration) have empirically further demonstrated beyond doubt that he essentially was correct and, more importantly, because no alternative theory to explain the same empirical evidences has been proposed by Wegener's opponents.

We simply highlight that Callebaut, et al. (2012, 2013) did not provide any better and/or alternative physical solar theory that explains why planetary harmonics characterize solar records at multiple time scales, including the 11-year solar cycle, as found by numerous authors. In De Jager and Duhan (2011) they just provided some highly qualitative and undemonstrated conjectures about the origin of the various secular and multisecular solar oscillations that, on the contrary, the planetary theory empirically predicts and hindcasts for millennia (e.g.: Scafetta, 2012c; Abreu et al., 2012). Callebaut et al. (2012) ruled out a planetary cause of solar variability by simply showing some (Newtonian classical physics) calculations (to explain the behavior of a non-Newtonian system such as the Sun) that just happen to be unable to demonstrate a physical link. We believe that Callebaut, de Jager and Duhau's physical argument and, in particular, the conclusion they derive from it are inadequate in front of the empirical evidences; their result may simply imply that the physical problem needs to be addressed in a different way than what they have proposed.

Perhaps, Scafetta's approach (2012d) is more appropriate; perhaps improved and extended physical theories may be proposed in the future.

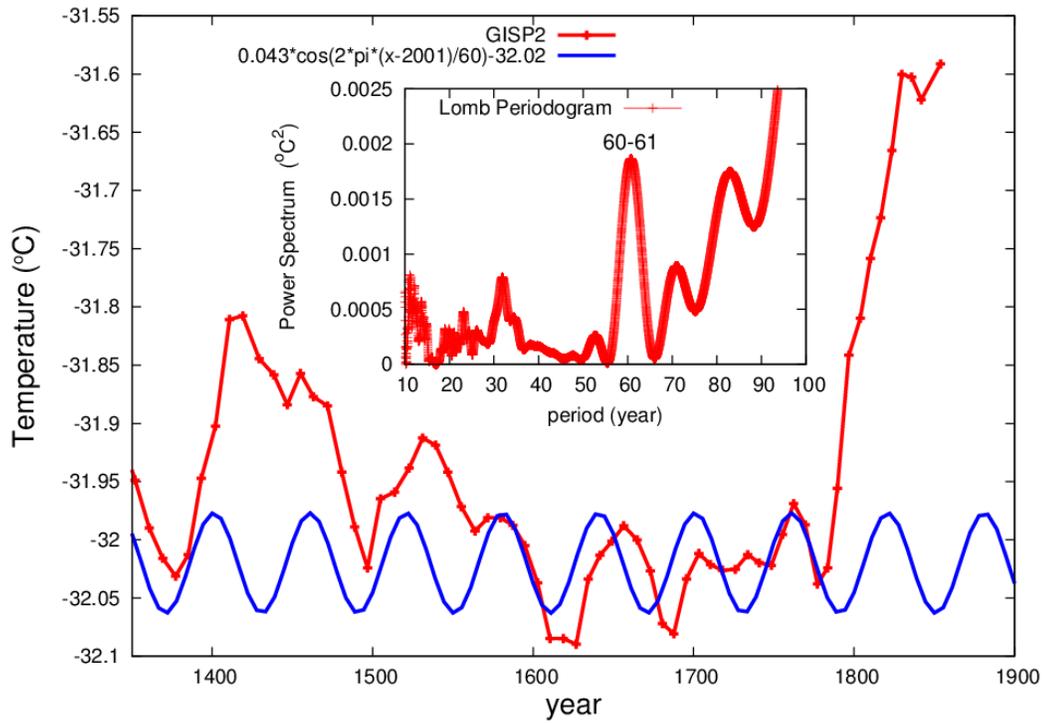

**Figure 1.** The GISP2 record in Humlum et al. (2011) from 1350 AD to 1850 AD (red), its power spectrum evaluation (in the insert) and the 60-year astronomical cycle proposed by Scafetta (2010) (blue). Despite a possible temporal and amplitude uncertainty, the GISP2 record presents a quasi 60-year modulation (clearly revealed in the power spectrum 60-61 yr peak) in relatively good phase with the 60-year astronomical harmonic proposed by Scafetta (2010). There appears to be a small time-lag perhaps due to a climatic heat capacity delay or to a possible proxy timing error (Bender et al., 1997).